The Great Oxidation Event (GOE):
Biogeochemical Feedback and Tipping Points

Andrew P. Ingersoll
Division of Geological and Planetary Sciences
California Institute of Technology
Pasadena, CA 91125
api@caltech.edu

Abstract

Approximately 1.4 Ga after life first appeared, atmospheric oxygen suddenly jumped by more than an order of magnitude over a 20-50 Ma period. The contrast between these two timescales does not seem to be due to any sudden, large amplitude change in external forcing. However, it could be due to processes intrinsic to the geobiological system itself, namely, positive feedback between atmospheric oxygen and photosynthetic bacteria: More oxygen leads to more photosynthesis, which leads to more oxygen, and so on. Already-published feedbacks include buildup of an ozone shield and nutrient production by oxidative weathering. The feedback proposed here is the 15-fold greater efficiency of aerobic vs anaerobic respiration and the tight coupling of respiration and photosynthesis inside the cell. As in the climate system, feedback leads to tipping points, where a rapid, large amplitude change in the state of the system occurs. For the geobiological system, the GOE is the tipping point, and the long buildup before the GOE is the gradual oxidation of the crust and ocean, due either to burial of organic matter, oxidation of volcanic gases, or escape of hydrogen to space. The feedback hypothesis is a framework for interpreting observations leading to the GOE.

The most precise timing of the GOE comes from the isotopes of sulfur (Farquhar et al., 2000) in sedimentary rocks. Observations place the event 2.33 Ga ago, 1.4 Ga after life first appeared (Blankenship, 1992; Bosak et al., 2013; Nutman et al., 2016), and its duration at ≤ 30 Ma (Luo et al., 2016). The time span of the duration is limited by the precision of the observations and may be just an upper bound. Evidence for the GOE is that departures of the isotopes of sulfur from a global standard suddenly started following a linear relation with respect to mass. The measurement (Farquhar et al., 2011) is illustrated in Fig. 1. The interpretation is that instead of appearing in a variety of valence states, as happens for volcanic gases in the atmosphere, most of the sulfur around the globe was suddenly oxidized to sulfate, $SO_4^{2-}$, which led to mass-dependent fractionation.

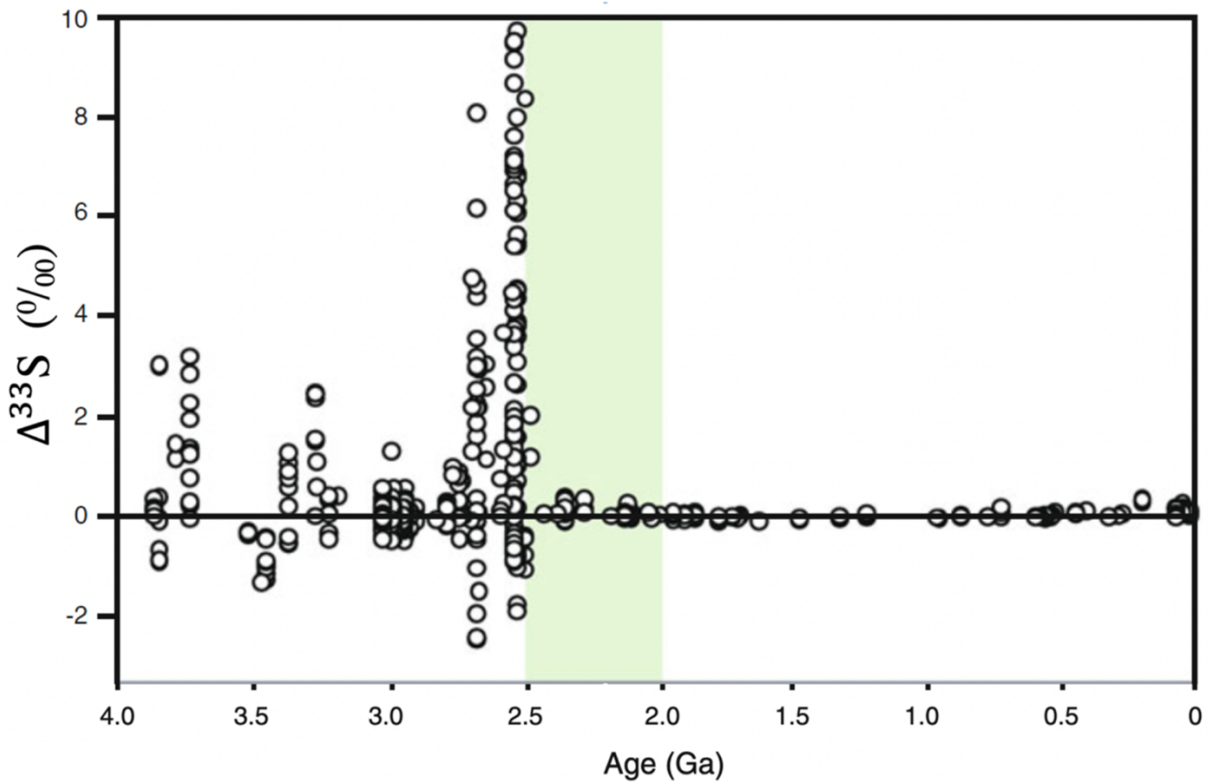

Fig. 1. The transition from mass-independent fraction to mass-dependent fractionation of sulfur isotopes in sediments from around the world. Radioactive elements in the sediment provide the age. The three isotopes $^{32}S$, $^{33}S$, and $^{34}S$ provide the amount of nonlinearity with respect to mass, $\Delta^{33}S$, which is the amount of mass-independent fractionation. From (Farquhar et al., 2011).

There are two fundamental questions. What caused the long (1.4 Ga) delay in the appearance of $O_2$ after life first appeared? And why did the delay end so abruptly? The oxygen levels during the long delay are highly uncertain (Lyons et al., 2014; Catling and Zahnle, 2020). The delay is often attributed to reduced gases like $H_2S$, $SO_2$, $CH_4$ and $H_2$, which are emitted by volcanoes and would tend to maintain anoxic conditions at the surface, i.e., in the crust, atmosphere and oceans (Holland, 2002). The earliest life forms were leaving their mark on the isotopes of carbon in sediments dating at least to 3400 Ga ago, where negative values of $\delta^{13}C$ in organic matter are clearly distinguishable from zero values in carbonates (Hayes and Waldbauer, 2006). These life forms did not necessarily produce $O_2$. In anaerobic environments, they could have used anoxygenic photosynthesis, still turning the energy of sunlight into chemical energy in the form of sugars. Subduction of the resulting organic matter into the deep mantle could have helped oxidize the upper mantle and crust by leading to a separation of carbon into reduced forms like graphite and diamond and oxidized forms like carbonate and $CO_2$. The oxidized forms could feed into the mixture of volcanic gases and would tend to weaken their reducing power (Duncan and Dasgupta, 2017). Escape of hydrogen to space would also weaken the reducing power of the volcanic gases if the crust, upper mantle, ocean, and atmosphere were behaving as a single system (Catling et al., 2001; Zahnle et al., 2013; Kelemen and Manning, 2015). Oxidation of the surface, including the crust and upper mantle, would have been slow, simply because it is a large reservoir, and that may have caused the long delay. Further delay might have been caused by the limited availability of nutrients, particularly nitrate and phosphorus (Rambler and Margulis, 1976; Fennel et al., 2005; Olejarz et al., 2021; Alcott et al., 2022). Modern oxygen reservoirs, like $O_2$ in the atmosphere, sulfate in the oceans, and oxidized iron in the crust and upper mantle, are maintained on a much shorter timescale by biological activity.

The other question, having to do with the abrupt jump in atmospheric $O_2$, has been attributed to $O_2$ crossing a threshold that arises due to feedback - more oxygen in the atmosphere makes the bacteria produce even more oxygen, leading to a transition to a new stable state that is substantially different from the old one. In dynamical systems, such jumps are called saddle node bifurcations. In the climate system, they are called tipping points. Ice-albedo feedback is an example – the colder it gets, the more Earth's surface is covered with ice, and the more sunlight is reflected to space, making the Earth even colder (Budyko, 1969). Another example is water vapor feedback – the warmer it gets, the more water vapor is in the atmosphere, and the more heat is trapped since water vapor is a greenhouse gas (Ingersoll, 1969).

Figure 2 shows an idealized model of ice-albedo feedback. The equation is

$$\frac{dT}{dt} = \frac{\sigma}{C}\left[T_{eq}^4(1 - A(T)) - \epsilon T^4\right] \qquad (1)$$

Here $T$ is the surface temperature, assumed constant over the area of the planet, and $T_{eq}$ is the temperature of a perfectly absorbing, black isothermal sphere in equilibrium with sunlight at Earth orbit. The Stefan-Boltzmann constant is $\sigma$; the heat capacity is $C$ in units of energy per unit surface area per K; and $[1 - A(T)]$ is the fraction of sunlight that is absorbed. It represents the feedback, because it has the output $T$ affecting the input, which is the rate of change of $T$. Finally, $\epsilon$ is the infrared emissivity, a constant independent of $T$ because we are not considering water vapor feedback in this example.

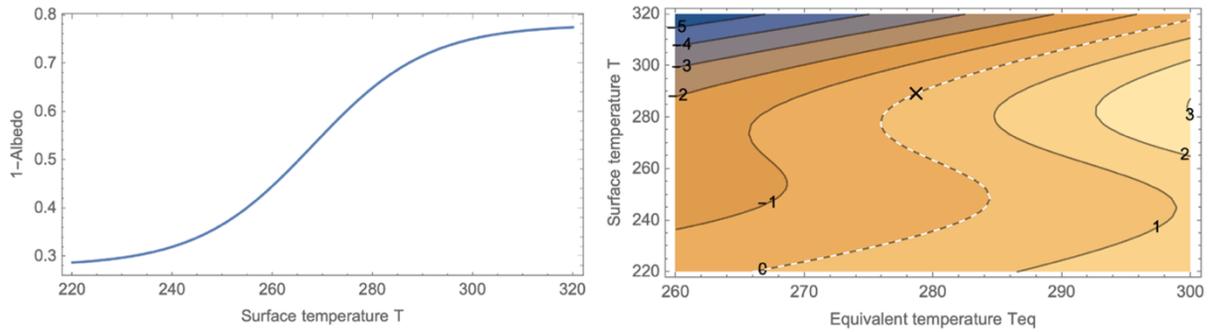

Fig. 2 (left). Fraction (1 - A) of incident sunlight that is absorbed, where A is the visual albedo. (1 - A) increases from 0.28 at low T to 0.78 at high T, creating a positive feedback in the climate system. Fig. 2 (right). Contour plot of the rate of change dT/dt - the right side of Eq. (1) - as a function of surface temperature $T$ (y-axis) and solar forcing $T_{eq}$ (x-axis), the latter being the temperature of a perfectly absorbing, black isothermal sphere in equilibrium with sunlight. The units of dT/dt depend on the size of the thermal reservoir and are arbitrary in this figure. Positive contours are where dT/dt > 0, and negative contours are where dT/dt < 0. The zero contour, which is dashed, shows the equilibrium solutions, where T is not changing, and they group into three parts defined by the slope. Near the upper and lower parts, dT/dt is toward the zero contour, indicating that those parts are stable. Near the middle part, dT/dt is away from the zero contour - positive above it and negative below it, indicating that that part is unstable. The left and right extrema of the zero contour are the tipping points, and they define the places where the system can jump from one stable equilibrium to the other. The X at the center denotes the present Earth according to this model. It is a stable, warm-Earth solution. The lower part of the zero contour is a stable, snowball-Earth solution. A decrease of $T_{eq}$ would push the warm Earth to the lefthand tipping point, causing transition to a snowball Earth. An increase of $T_{eq}$ would push the snowball Earth toward the righthand tipping point, causing transition to a warm Earth.

Several types of feedback have been proposed to explain the GOE. Multiple equilibria, with sudden jumps from one to the other, are a common feature. We first discuss the different feedback mechanisms that have been proposed, and then we propose a new mechanism. Multiple equilibria – a bistatic system – could arise (Goldblatt et al., 2006; Gregory et al., 2021; Wogan et al., 2022) where one steady state exists at $O_2$ partial pressures of $10^{-3}$ mbar or less and is held there by oxidation of $CH_4$, and another steady state exists at ~1 mbar or greater and is held there by an ozone shield. The ozone shield not only protects the bacteria from cell damage, it also protects water vapor from photodissociation. The latter is important because photodissociation leads to production of hydroxyl, and hydroxyl speeds up destruction of $O_2$ by oxidation of $CH_4$ (Goldblatt et al., 2006). A common feature of these models is the existence of two stable equilibria - a low-oxygen state and a high-oxygen state - with a threshold for jumping rapidly from one to another, i.e., with time scales between $10^2$ and $10^5$ years (Wogan et al., 2022).

Other feedbacks involve the abundance of nutrients - phosphorous in particular - and the abundance of reducing substances like FeO, $H_2S$, $CH_4$ and $H_2$. When the former are sufficiently high and the latter are sufficiently low, a threshold is reached where oxygenic photosynthesis is favored over anoxygenic photosynthesis. The transition can be either gradual and reversible or sudden and irreversible, depending on sources and sinks of oxygen (Olejarz et al., 2021). Extensive phosphorous recycling combined with a general increase in the input of phosphorous due to continental weathering could have the same effect (Alcott et al., 2022). Others find positive feedback, in which low $O_2$ leads to ferrous phases of iron, which are more efficient sinks of phosphate than ferric iron, leading to lower amounts of $O_2$ (Laakso and Schrag, 2017). These models all have geophysical processes controlling the time scale.

Another study (Fakhraee and Planavsky, 2024) uses a dynamical system approach to fit $dp(O_2)/dt$ from before the GOE to the present. Rather than accept the parameter values of published models, the authors vary them in a stochastic fashion to fit not only the long-term trends but also the fluctuations, pulses, and large swings of atmospheric oxygen that are in the geologic record. They find three kinds of stable equilibrium - at low, medium, and high $p(O_2)$, which are typical of the Archean, Proterozoic, and modern eras, and they identify the unstable equilibria and tipping points in between.

We propose a new and different feedback mechanism, which arises from the 15-times greater efficiency of aerobic to anaerobic respiration and the tight coupling of photosynthesis and respiration. This is strong positive feedback – the more oxygen is in the atmosphere, the more rapidly the bacteria produce oxygen. That feedback is our model for the GOE. It is a tipping point – an abrupt change in the state of a system – that would have taken place on a short, biological time scale. In contrast, oxidizing the surface would have taken place on a long, geological time scale. The US Geological Survey's upper limit for anoxic water has a partial pressure of 1.2 mbar at equilibrium, corresponding at 25 C° to a solution with 50 $\mu g L^{-1}$, or ~1.5 $\mu$M of dissolved $O_2$ (Wikipedia entries: Henry's Law and Anoxic Waters). This upper limit is also known as the Pasteur point, where microorganisms adapt from anaerobic respiration to aerobic respiration. The partial pressure of $O_2$ in the present atmosphere is 209.5 mbar.

Respiration is the process by which organisms use chemical energy to do work. It is measured in the number of ATP molecules per atom of carbon. The chemical reaction most relevant to our feedback mechanism is glucose plus oxygen yields $CO_2$ plus water. That chemical reaction is the opposite of photosynthesis, which captures the energy of sunlight to make glucose. If glucose is represented symbolically as $CH_2O$, the reaction requires $O_2$ as the electron acceptor. It is the textbook example of aerobic respiration and yields ~30 ATP molecules per carbon atom (Flurkey, 2010). In anaerobic respiration, the electron acceptor may be the oxygen in oxidized iron ($Fe_2O_3$), nitrate ($NO_3^-$), sulfate ($SO_4^{-2}$) or carbon dioxide. There are many reactions and many ways to obtain energy without $O_2$, including fermentation and methanogenesis. The former produces lactic acid and ethanol from glucose (Stal and Moezelaar, 1997), yielding 2 ATP molecules per carbon atom. Methanogenesis produces similar or smaller amounts of energy (Buan, 2018; Lyu et al., 2018). An example is $CO_2$ plus $4H_2$ making $CH_4$ and $2H_2O$. The $H_2$ either comes from decaying organic matter (e.g., in ruminants) or from water-rock interaction (oxidation of iron).

In cyanobacteria today, respiration and photosynthesis are tightly coupled. They share the same metabolic pathway, the thylakoid membrane, for electron transfer (Kirchhoff, 2014). The molecule that carries the electrons, plastoquinone, functions as a redox sensor that drives the magnitude and direction of the transfer to maintain metabolic and redox homeostasis (Havaux, 2020; Shimakawa et al., 2021). Respiration supplies metabolic energy to the cell during the night, and it protects the cell from excess oxygen during the day. The close relationship also occurs in mm-thick mats and films, where the embedded heterotrophs, organisms that do not engage in photosynthesis, provide respiration that enables photosynthesis by the photoautotrophs, the cyanobacteria, that do (Kuhl et al., 1996; Ploug, 2008; Staal et al., 2003). Finally, the fact that about half of global primary production (GPP) is used for respiration by the photoautotrophs, and the other half, the net primary production (NPP) goes out to feed the heterotrophs (Field et al., 1998), is evidence that photosynthesis and respiration are coupled on a global scale as well (Paumann et al., 2005).

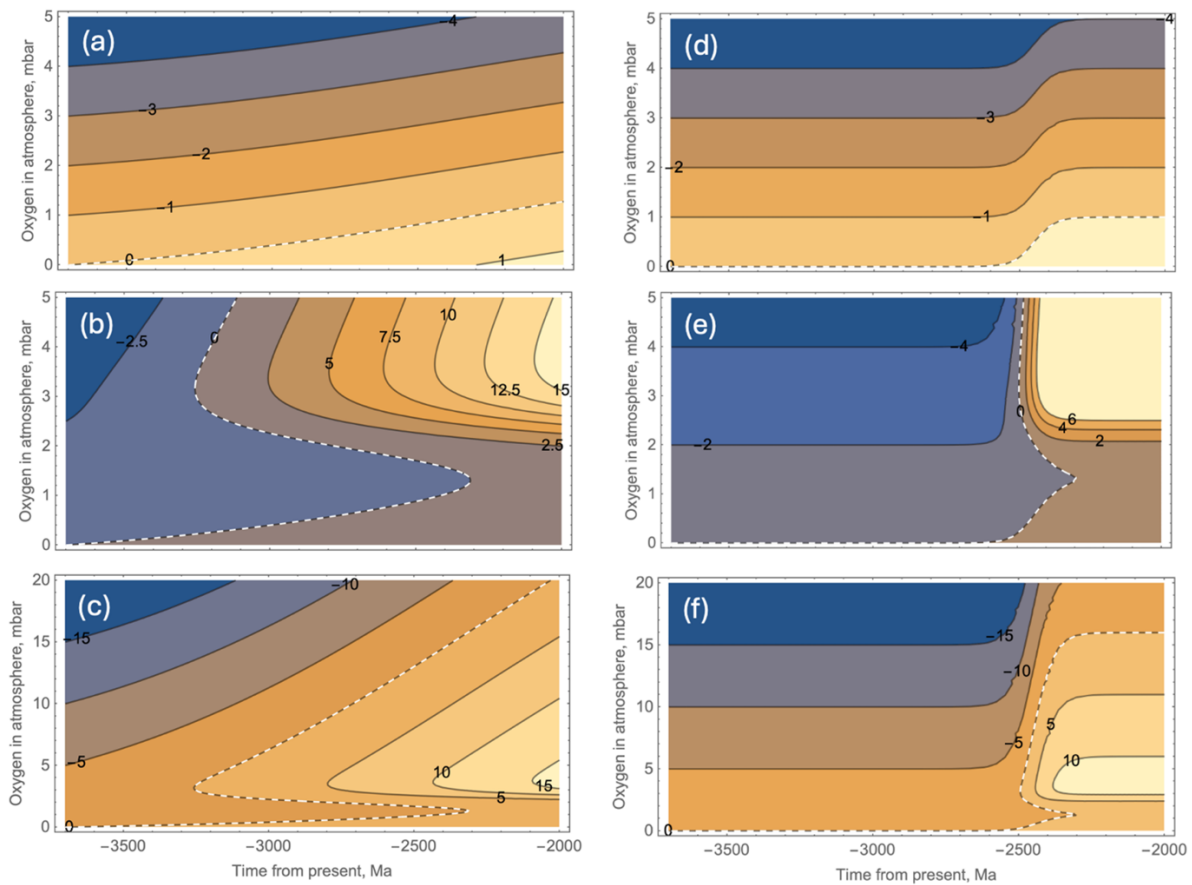

Fig. 3. The GOE represented as a tipping point that occurs when the atmosphere and surface oceans cross the Pasteur point from anoxic to aerobic conditions, about 1.2 mbar partial pressure or a 50 μM solution. One version of the model (a, b, c) has $P_{volc}(t)$ with a steady increase of anoxic volcanic gases - the dotted line of panel (a) - and a ~15-fold jump in pressure at -2.3 Ga (b and c). The other version shows $P_{volc}(t)$ with a long period of anoxic volcanic gases (d) followed by an increase beginning at about 2.6 Ga and a ~15-fold increase in pressure at -2.3 Ga (e and f). The dashed line is where the system is at equilibrium. Contours to the left and above the line (negative numbers) have pressure decreasing. Contours to the right and below the line have pressure increasing. The tipping point, where the jump in $P_{atm}$ occurs, is at t = -2.3 Ga.

Figure 3 shows a conceptual model for the growth of atmospheric O₂ over time. The structure of the model is like that in Fig. 2, except the feedback is due to photosynthesis being coupled to respiration, which is 15 times more efficient above the Pasteur point than below it. In contrast, the feedback in Eq. (1), i.e., the (1 – A) term, only increases the effect of solar forcing by a factor of ~3. We define $P_{atm}$ as the partial pressure of O₂ in the atmosphere and $P_{volc}$ as the effect of volcanic gases - the $P_{atm}$ that the atmosphere would have in steady state if feedback were absent. We choose $P_{volc}(t)$ as input to the model. It is the dashed line in panels (a) and (d).

The equation for atmospheric O₂, from which Fig. 3 is derived, is

$$\frac{dP_{atm}}{dt} = \frac{1}{\tau}[P_{volc}\{1 + f_{bck}(P_{atm})\} - P_{atm}] \tag{2}$$

Feedback $f_{bck}$ amplifies the effect of $P_{volc}$ by adding another term to $dP_{atm}/dt$. The feedback is an increasing function of $P_{atm}$, which leads to the multiple equilibria and tipping points. We assume the tight coupling between photosynthesis and respiration applies to both low and high values of $P_{atm}$.

The parallels between Eqs. (1) and (2) are instructive. The input $P_{volc}$ is analogous to the solar forcing. It is related to the oxidation state - fugacity - of the volcanic gases. The output $P_{atm}$ is analogous to the temperature. It is the partial pressure of O₂ in the atmosphere. The units are mbar. The feedback $f_{bck}$ is a dimensionless quantity analogous to (1 - A) in Eq. (1). Like (1 - A), it is an increasing function of the output, which is $P_{atm}$ in mbar. The magnitude of the feedback is << 1 when $P_{atm} << P_{ppt}$, and it rises to ~15 when $P_{atm} >> P_{ppt}$, where $P_{ppt}$ is the Pasteur point, ~1.2 mbar according to the USGS. The functional form has no physical basis and is used to help visualize the qualitative concepts developed in this model. It has been tuned to make the tipping point coincide with the Pasteur point at a time 2.3 Ga before the present.

$$f_{bck}(P_{atm}) = 15\{1 + tanh[4(0.4P_{atm} - 1)]\}/2 \tag{3}$$

Finally, the time constant $\tau$ is related to the total plant biomass (Bar-On et al., 2018) divided by the net primary production NPP (Field et al., 1998), which is the total photosynthesis rate minus that used by the photoautotrophs in their respiration. The modern value of $\tau$ is ~5000 years, but it could have been much longer in the past when climate and nutrients were different and the NPP was less.

A literature search finds no papers that use the increased efficiency of respiration and the tight coupling of photosynthesis and respiration to provide positive feedback during the GOE. The increased efficiency of respiration at O₂ partial pressures above the Pasteur point is well established. The tight coupling of respiration and photosynthesis is also well established. It even has a name – photorespiration, where plants oxidize the glucose as soon as it is produced, deliberately wasting energy. The plants do it to avoid cell damage from excessive sunlight and oxidative stress. One group (Wogan et al., 2022) mentions aerobic respiration, but they describe it as one of the several biological feedbacks that their model does not consider (see their discussion section). Biological feedback immediately *after* the GOE is widely studied (e.g., Schopf, 2014). A model that uses the tight coupling of respiration and photosynthesis (Falkowski and Godfrey, 2008) does so in the context of events *after* the GEO, with massive gene transfer to eukaryotic host cells as bacteria evolve to compete in their new environment.

A problem with the oxygen feedback mechanism illustrated in Fig. 3 is that it should not exist at $O_2$ partial pressures below 1.2 mbar, because aerobic respiration should not exist there. The tight coupling implies that aerobic photosynthesis should not exist at those pressures either. However, Figs. 1-3 are one-dimensional. Fig. 1 shows one worldwide distribution of sulfur isotopes as a function of time and ignores geographic variability. Fig. 2 shows one temperature and one albedo for the whole Earth, and ignores differences between poles and equator, continents and oceans, and cloudy and cloud-free areas. Fig. 3 shows one partial pressure of $O_2$ in the atmosphere, ignoring sampling bias and geographic variability. But, if there were places on Earth - niches - where $O_2$ partial pressures were greater than 1.2 mbar, even though the whole atmospheric partial pressure were lower, then the feedback could occur in those places.

Stromatolites, which are mineral microbial mats, are one such candidate. The mats are colonies of organisms, mostly bacteria and archaea, that form biofilms that cement mineral grains, mostly carbonate, together (Bosak et al., 2013). They were ubiquitous before the GOE. They are among the earliest evidence of life on Earth and are still active today. Stromatolites thrive in bodies of water and wetlands and vary in size from a few cm to tens of m. They come in a variety of shapes including flat, rounded, domical and tufted (Flannery and Walter, 2012). The laminae of the mats are cm in thickness and have sub-mm bubbles or holes (fenestrate) embedded in them (Bosak et al., 2009; Wilmeth et al., 2019). Laboratory experiments with modern microbial mats show that $O_2$ is produced within the bubbles by photosynthesis (Bosak et al., 2010). Bubbles of $O_2$ tend to rise as they replace $CO_2$ due to the higher molecular weight of the latter. They do not explode, since volume and pressure before and after are equal. Ultimately, the contents of the bubbles will diffuse into the outside atmosphere. The important point is that oxygenic photosynthesis could have operated inside the stromatolite bubbles with arbitrarily low partial pressure of $O_2$ outside, despite the tight coupling of photosynthesis and respiration in both places.

Estimating the partial pressure of $O_2$ in the Archean atmosphere is difficult: Different studies have come to different conclusions. Carbon in minerals as old as 3.8 Ga are often depleted in the heavier isotope $^{13}C$, which is indicative of biological activity (Mojzsis et al., 1996). The $^{13}\delta C$ value has remained negative and relatively constant over the age of the Earth (Hayes and Waldbauer, 2006). Presumably the bacteria were making glucose, but this does not mean they were producing $O_2$. Anoxygenic photosynthesis might have used ferrous iron or sulfide as the electron donor for fixing $CO_2$ (Blankenship, 1992; Olson and Blankenship, 2004; Crowe et al., 2008; Hamilton, 2019). One study concludes that $O_2$ was widespread and pervasive in microaerobic marine environments hundreds of Ma before the GOE (Waldbauer et al., 2011). But other studies show that the cited evidence, e.g., trace metal oxidation of sediments, including sulfur, might be due to oxidative weathering *after* the GOE (Slotznick et al., 2022). At modern rates, oxygenic photosynthesis would have overwhelmed Earth's redox buffers in ~100 kyr, suggesting to one group (Ward et al., 2016) that oxygenic photosynthesis arose shortly before the rise of oxygen, not hundreds of million years before it. Of course, modern rates might not apply before the GOE.

A comprehensive review (Ostrander et al., 2021) of inorganic indicators of oxygen concludes that there was not only one oxidation event around 2.5 Ga ago, 200Ma before the GOE, but also 2 or more earlier Archean oxidation events dating back to 3.5 Ga before the present. The inorganic indicators include both redox-sensitive elements like Mo and Re and elements that undergo isotopic fractionation. Oxidative weathering of the continents delivers the $MoO_4^{2-}$ and the $ReO_4^-$ to the ocean when $O_2$ is present (Anbar et al., 2007). And a disparate array of elements including N, Mo, Tl, U, and Se give information about marine oxidation through their isotope ratios (Ostrander et al., 2021). The important point is that there were several distinct events: The 2.5 Ga oxygenation, also termed a "whiff" of oxygen (Anbar et al., 2023), was more than just a ramp-up to the GOE.

Inevitably, the sediments used in these studies come from a small number of locations, mostly western Australia and South Africa, so there might be differences of weathering across the GOE boundary, causing the apparent timescale to thicken. This is called the crustal memory effect. However, the $\Delta^{33}S$ values from post-GOE deposits worldwide vary between $\pm 0.3$ ‰, which is an order of magnitude smaller than the pre-GOE variation (see Fig. 1). The one instance of sediments with $\Delta^{33}S > 3$‰ getting mixed into an overlying stratum with $\Delta^{33}S < 0.3$‰ involved a single site with a layer thickness of < 1m and a time difference across the layer of only < $10^5$ years (Uveges et al., 2023) and could be due to an atmospheric dynamics effect.

The genomes of cyanobacteria tell the story through a different medium - the relative rates of evolution. Cyanobacteria are the only bacteria capable of oxygenic photosynthesis. One study (Shih et al., 2017) places an upper limit of 2.5-2.6 Ga before the present for the origin of oxygenic photosynthesis. A related study (Soo et al., 2017) argues that all three classes of cyanobacteria acquired aerobic respiratory complexes after the GOE, strengthening the possibility that the GOE was directly caused by the evolution of cyanobacteria. However, another genomic analysis (Fournier et al., 2021) concludes that oxygenic photosynthesis evolved at least several hundred million years before the GOE. The last study argues that relative dating information from horizontal gene transfers between cyanobacterial lineages greatly improves the precision of the age estimates. A study of phenotypical traits of early cyanobacteria finds a sudden increase in the rate of evolutionary innovation immediately after the GOE but not before it (Uyeda et al., 2016). This would imply that cyanobacterial evolution was not a cause of the GOE but instead was a reaction to it.

The tight coupling of photosynthesis and respiration affects the chemical balances that could have occurred before the GOE. Reduced volcanic gases have traditionally been cited for the long delay between the appearance of bacteria and the sudden rise of oxygen (Catling et al., 2001; Kasting, 2001; Holland, 2002). However, if oxygenic photosynthesis were occurring, the reduced volcanic gases could not be maintaining atmospheric $O_2$ below the Pasteur point, because then oxygenic photosynthesis would shut down. Instead, anoxygenic photosynthesis would make organic matter using ferrous iron, sulfide, methane, or hydrogen as electron donors (Crowe et al., 2008; Hamilton, 2019; Olson, 2006), producing ferric iron, sulfate, and carbon dioxide. During subduction, separation of organic matter into oxidized carbon and reduced carbon and burial of the latter as graphite and/or diamond at depths greater than 250 km, i.e., in the deep mantle, would have slowly oxidized the crust and upper mantle (Duncan and Dasgupta, 2017; Hayes and Waldbauer, 2006). This would have weakened the reducing power of the volcanic gases, since they presumably are created in the crust and upper mantle.

Oxidation of methane followed by escape of hydrogen to space (Catling et al., 2001; Zahnle et al., 2013; Zahnle et al., 2019) would either have directly oxidized the methane resulting from decaying organic matter or oxidized the iron and sulfide in the crust and upper mantle. Like the burial of graphite and diamond, the net effect would have been to weaken the reducing power of the volcanic gases. Burial of organic matter followed by oxidation would have been slow because less organic matter, perhaps orders of magnitude less, was being made during the Archean compared to the present. In either case, the reducing power of the volcanic gases might have gotten so low, and the partial pressure of $O_2$ might have gotten so high that one of the tipping points mentioned earlier was reached.

It is interesting to treat the crust, upper mantle, oceans and atmosphere as a single system, its parts linked together by subduction and volcanic outgassing. The upper mantle, which defines the lower part of the lithosphere, extends down to 50-300 km. An estimate based on the modern carbon cycle (Kelemen and Manning, 2015) suggests that this system is almost closed on a 5-10 Ma time scale, which is surprisingly short. Transport of diamonds into the lower mantle is a small fraction of the global carbon inventory according to this study. The 5-10 Ma time scale is probably shorter than the duration of the GOE, to the extent that it can be resolved in the sulfur isotope record (Fig. 1). The rate of hydrogen escape is uncertain by one or two orders of magnitude (Zahnle et al., 2019), but it would probably be negligible when integrated over a 5-10 Ma interval. Even if the system behaved as if it were completely closed on this time scale, fluctuations in exchange rates between different parts of the system could lead to multiple oxidation events (Ostrander et al., 2021). If the time scale is correct, the 1.4 Ga delay should not be regarded as the slow filling of a large geologic reservoir but rather as a succession of short-term events whose integral slowly evolved to allow aerobic respiration and oxygenic photosynthesis to occur.

Acknowledgement

This research was supported by NASA under grant/cooperative agreement number 80NSSC20K0555 and by Caltech under Provost Academic Programs.

Conflict of Interest: None

References

Alcott, L.J., Mills, B.J.W., Bekker, A., Poulton, S.W., 2022. Earth's Great Oxidation Event facilitated by the rise of sedimentary phosphorus recycling. Nat. Geosci. 15, 210-+. https://doi.org/10.1038/s41561-022-00906-5

Anbar, A.D., Duan, Y., Lyons, T.W., Arnold, G.L., Kendall, B., Creaser, R.A., Kaufman, A.J., Gordon, G.W., Scott, C., Garvin, J., Buick, R., 2007. A whiff of oxygen before the Great Oxidation Event? Science 317, 1903–1906. https://doi.org/10.1126/science.1140325

Anbar, A.D., Buick, R., Gordon, G.W., Johnson, A.C., Kendall, B., Lyons, T.W., Ostrander, C.M., Planavsky, N.J., Reinhard, C.T., Stueken, E.E., 2023. Reexamination of 2.5-Ga whiff' of oxygen interval points to anoxic ocean before GOE. Sci. Adv. 9, eabq3736. https://doi.org/10.1126/sciadv.abq3736

Bar-On, Y.M., Phillips, R., Milo, R., 2018. The biomass distribution on Earth. Proc. Natl. Acad. Sci. U. S. A. 115, 6506–6511. https://doi.org/10.1073/pnas.1711842115

Blankenship, R., 1992. Origin and Early Evolution of Photosynthesis. Photosynth. Res. 33, 91–111. https://doi.org/10.1007/BF00039173

Bosak, T., Bush, J.W.M., Flynn, M.R., Liang, B., Ono, S., Petroff, A.P., Sim, M.S., 2010. Formation and stability of oxygen-rich bubbles that shape photosynthetic mats. Geobiology 8, 45–55. https://doi.org/10.1111/j.1472-4669.2009.00227.x

Bosak, T., Knoll, A.H., Petroff, A.P., 2013. The meaning of stromatolites, in: Jeanloz, R. (Ed.), Annual Review of Earth and Planetary Sciences, Vol. 41. Annual Reviews, Palo Alto, pp. 21–44. https://doi.org/10.1146/annurev-earth-042711-105327

Bosak, T., Liang, B., Sim, M.S., Petroff, A.P., 2009. Morphological record of oxygenic photosynthesis in conical stromatolites. Proc. Natl. Acad. Sci. U. S. A. 106, 10939–10943. https://doi.org/10.1073/pnas.0900885106

Buan, N.R., 2018. Methanogens: pushing the boundaries of biology. Emerg. Top. Life Sci. 2, 629–646. https://doi.org/10.1042/ETLS20180031

Budyko, M., 1969. Effect of solar radiation variations on climate of Earth. Tellus 21, 611-. https://doi.org/10.3402/tellusa.v21i5.10109

Catling, C., Zahnle, K., 2020. The Archean atmosphere. Sci. Adv. 6, eaax1420. https://doi.org/10.1126/sciadv.aaX1420

Catling, D.C., Zahnle, K.J., McKay, C.P., 2001. Biogenic methane, hydrogen escape, and the irreversible oxidation of early Earth. Science 293, 839–843. https://doi.org/10.1126/science.1061976

Crowe, S.A., Jones, C., Katsev, S., Magen, C., O'Neill, A.H., Sturm, A., Canfield, D.E., Haffner, G.D., Mucci, A., Sundby, B., Fowle, D.A., 2008. Photoferrotrophs thrive in an Archean Ocean analogue. Proc. Natl. Acad. Sci. U. S. A. 105, 15938–15943. https://doi.org/10.1073/pnas.0805313105

Duncan, M.S., Dasgupta, R., 2017. Rise of Earth's atmospheric oxygen controlled by efficient subduction of organic carbon. Nat. Geosci. 10, 387-+. https://doi.org/10.1038/NGEO2939

Fakhraee, M., Planavsky, N., 2024. Insights from a dynamical system approach into the history of atmospheric oxygenation. Nat. Commun. 15, 6794. https://doi.org/10.1038/s41467-024-51042-0

Falkowski, P.G., Godfrey, L.V., 2008. Electrons, life and the evolution of Earth's oxygen cycle. Philos. Trans. R. Soc. B-Biol. Sci. 363, 2705–2716. https://doi.org/10.1098/rstb.2008.0054

Farquhar, J., Bao, H.M., Thiemens, M., 2000. Atmospheric influence of Earth's earliest sulfur cycle. Science 289, 756–758. https://doi.org/10.1126/science.289.5480.756

Farquhar, J., Zerkle, A.L., Bekker, A., 2011. Geological constraints on the origin of oxygenic photosynthesis. Photosynth. Res. 107, 11–36. https://doi.org/10.1007/s11120-010-9594-0

Fennel, K., Follows, M., Falkowski, P.G., 2005. The co-evolution of the nitrogen, carbon and oxygen cycles in the Proterozoic ocean. Am. J. Sci. 305, 526–545. https://doi.org/10.2475/ajs.305.6-8.526

Field, C.B., Behrenfeld, M.J., Randerson, J.T., Falkowski, P., 1998. Primary production of the biosphere: Integrating terrestrial and oceanic components. Science 281, 237–240. https://doi.org/10.1126/science.281.5374.237

Flannery, D.T., Walter, M.R., 2012. Archean tufted microbial mats and the Great Oxidation Event: new insights into an ancient problem. Australian Journal of Earth Sciences 59, 1–11. https://doi.org/10.1080/08120099.2011.607849

Flurkey, W.H., 2010. Yield of ATP molecules per glucose molecule. J. Chem. Educ. 87, 271–271. https://doi.org/10.1021/ed800102g

Fournier, G.P., Moore, K.R., Rangel, L.T., Payette, J.G., Momper, L., Bosak, T., 2021. The Archean origin of oxygenic photosynthesis and extant cyanobacterial lineages. Proc. R. Soc. B-Biol. Sci. 288, 20210675. https://doi.org/10.1098/rspb.2021.0675

Goldblatt, C., Lenton, T.M., Watson, A.J., 2006. Bistability of atmospheric oxygen and the Great Oxidation. Nature 443, 683–686. https://doi.org/10.1038/nature05169

Gregory, B.S., Claire, M.W., Rugheimer, S., 2021. Photochemical modelling of atmospheric oxygen levels confirms two stable states. Earth Planet. Sci. Lett. 561, 116818. https://doi.org/10.1016/j.epsl.2021.116818

Hamilton, T.L., 2019. The trouble with oxygen: The ecophysiology of extant phototrophs and implications for the evolution of oxygenic photosynthesis. Free Radic. Biol. Med. 140, 233–249. https://doi.org/10.1016/j.freeradbiomed.2019.05.003

Havaux, M., 2020. Plastoquinone in and beyond photosynthesis. Trends Plant Sci. 25, 1252–1265. https://doi.org/10.1016/j.tplants.2020.06.011

Hayes, J.M., Waldbauer, J.R., 2006. The carbon cycle and associated redox processes through time. Philos. Trans. R. Soc. B-Biol. Sci. 361, 931–950. https://doi.org/10.1098/rstb.2006.1840

Holland, H.D., 2002. Volcanic gases, black smokers, and the Great Oxidation Event. Geochim. Cosmochim. Acta 66, 3811–3826. https://doi.org/10.1016/S0016-7037(02)00950-X

Ingersoll, A., 1969. Runaway greenhouse - A history of water on Venus. J. Atmos. Sci. 26, 1191–1198. https://doi.org/10.1175/1520-0469(1969)026<1191:TRGAHO>2.0.CO;2

Kasting, J.F., 2001. Earth history - The rise of atmospheric oxygen. Science 293, 819–820. https://doi.org/10.1126/science.1063811

Kelemen, P.B., Manning, C.E., 2015. Reevaluating carbon fluxes in subduction zones, what goes down, mostly comes up. Proc. Natl. Acad. Sci. U. S. A. 112, E3997–E4006. https://doi.org/10.1073/pnas.1507889112

Kirchhoff, H., 2014. Diffusion of molecules and macromolecules in thylakoid membranes. Biochimica et Biophysica Acta (BBA) - Bioenergetics, Dynamic and ultrastructure of bioenergetic membranes and their components 1837, 495–502. https://doi.org/10.1016/j.bbabio.2013.11.003

Kuhl, M., Glud, R.N., Ploug, H., Ramsing, N.B., 1996. Microenvironmental control of photosynthesis and photosynthesis-coupled respiration in an epilithic cyanobacterial biofilm. J. Phycol. 32, 799–812. https://doi.org/10.1111/j.0022-3646.1996.00799.x

Laakso, T.A., Schrag, D.P., 2017. A theory of atmospheric oxygen. Geobiology 15, 366–384. https://doi.org/10.1111/gbi.12230

Luo, G., Ono, S., Beukes, N.J., Wang, D.T., Xie, S., Summons, R.E., 2016. Rapid oxygenation of Earth's atmosphere 2.33 billion years ago. Sci. Adv. 2, e1600134. https://doi.org/10.1126/sciadv.1600134

Lyons, T.W., Reinhard, C.T., Planavsky, N.J., 2014. The rise of oxygen in Earth's early ocean and atmosphere. Nature 506, 307–315. https://doi.org/10.1038/nature13068

Lyu, Z., Shao, N., Akinyemi, T., Whitman, W.B., 2018. Methanogenesis. Curr. Biol. 28, R727–R732. https://doi.org/10.1016/j.cub.2018.05.021

Mojzsis, S.J., Arrhenius, G., McKeegan, K.D., Harrison, T.M., Nutman, A.P., Friend, C.R.L., 1996. Evidence for life on Earth before 3,800 million years ago. Nature 384, 55–59. https://doi.org/10.1038/384055a0

Nutman, A.P., Bennett, V.C., Friend, C.R.L., Van Kranendonk, M.J., Chivas, A.R., 2016. Rapid emergence of life shown by discovery of 3,700-million-year-old microbial structures. Nature 537, 535-+. https://doi.org/10.1038/nature19355

Olejarz, J., Iwasa, Y., Knoll, A.H., Nowak, M.A., 2021. The Great Oxygenation Event as a consequence of ecological dynamics modulated by planetary change. Nat Commun 12, 1–9. https://doi.org/10.1038/s41467-021-23286-7

Olson, J.M., 2006. Photosynthesis in the Archean Era. Photosynth. Res. 88, 109–117. https://doi.org/10.1007/s11120-006-9040-5

Olson, J.M., Blankenship, R.E., 2004. Thinking about the evolution of photosynthesis. Photosynth. Res. 80, 373–386. https://doi.org/10.1023/B:PRES.0000030457.06495.83

Ostrander, C.M., Johnson, A.C., Anbar, A.D., 2021. Earth's First Redox Revolution, in: Jeanloz, R., Freeman, K.H. (Eds.), ANNUAL REVIEW OF EARTH AND PLANETARY SCIENCES, VOL 49, 2021. Annual Reviews, Palo Alto, pp. 337–366. https://doi.org/10.1146/annurev-earth-072020-055249

Paumann, M., Regelsberger, G., Obinger, C., Peschek, G.A., 2005. The bioenergetic role of dioxygen and the terminal oxidase(s) in cyanobacteria. Biochim. Biophys. Acta-Bioenerg. 1707, 231–253. https://doi.org/10.1016/j.bbabio.2004.12.007

Ploug, H., 2008. Cyanobacterial surface blooms formed by Aphanizomenon sp and Nodularia spumigena in the Baltic Sea: Small-scale fluxes, pH, and oxygen microenvironments. Limnol. Oceanogr. 53, 914–921. https://doi.org/10.4319/lo.2008.53.3.0914

Rambler, M., Margulis, L., 1976. Comment on Egamis Concept of Evolution of Nitrate Respiration. Orig. Life Evol. Biosph. 7, 73–74. https://doi.org/10.1007/BF01218516

Schopf, J.W., 2014. Geological evidence of oxygenic photosynthesis and the biotic response to the 2400-2200 Ma "Great Oxidation Event." Biochem.-Moscow 79, 165–177. https://doi.org/10.1134/S0006297914030018

Shih, P.M., Hemp, J., Ward, L.M., Matzke, N.J., Fischer, W.W., 2017. Crown group Oxyphotobacteria postdate the rise of oxygen. Geobiology 15, 19–29. https://doi.org/10.1111/gbi.12200

Shimakawa, G., Kohara, A., Miyake, C., 2021. Characterization of light-enhanced respiration in cyanobacteria. Int. J. Mol. Sci. 22, 342. https://doi.org/10.3390/ijms22010342

Slotznick, S.P., Johnson, J.E., Rasmussen, B., Raub, T.D., Webb, S.M., Zi, J.-W., Kirschvink, J.L., Fischer, W.W., 2022. Reexamination of 2.5-Ga "whiff" of oxygen interval points to anoxic ocean before GOE. Sci. Adv. 8, eabj7190. https://doi.org/10.1126/sciadv.abj7190

Soo, R.M., Hemp, J., Parks, D.H., Fischer, W.W., Hugenholtz, P., 2017. On the origins of oxygenic photosynthesis and aerobic respiration in Cyanobacteria. Science 355, 1436–1439. https://doi.org/10.1126/science.aal3794

Staal, M., Hekkert, S.T.L., Harren, F.J.M., Stal, L.J., 2003. Effects of O-2 on N-2 fixation in heterocystous cyanobacteria from the Baltic Sea. Aquat. Microb. Ecol. 33, 261–270. https://doi.org/10.3354/ame033261

Stal, L.J., Moezelaar, R., 1997. Fermentation in cyanobacteria. Fems Microbiol. Rev. 21, 179–211. https://doi.org/10.1016/S0168-6445(97)00056-9

Uveges, B.T., Izon, G., Ono, S., Beukes, N.J., Summons, R.E., 2023. Reconciling discrepant minor sulfur isotope records of the Great Oxidation Event. Nat Commun 14, 279. https://doi.org/10.1038/s41467-023-35820-w

Uyeda, J.C., Harmon, L.J., Blank, C.E., 2016. A Comprehensive Study of Cyanobacterial Morphological and Ecological Evolutionary Dynamics through Deep Geologic Time. PLoS One 11, e0162539. https://doi.org/10.1371/journal.pone.0162539

Waldbauer, J.R., Newman, D.K., Summons, R.E., 2011. Microaerobic steroid biosynthesis and the molecular fossil record of Archean life. Proc. Natl. Acad. Sci. U. S. A. 108, 13409–13414. https://doi.org/10.1073/pnas.1104160108

Ward, L.M., Kirschvink, J.L., Fischer, W.W., 2016. Timescales of Oxygenation Following the Evolution of Oxygenic Photosynthesis. Orig. Life Evol. Biosph. 46, 51–65. https://doi.org/10.1007/s11084-015-9460-3

Wilmeth, D.T., Corsetti, F.A., Beukes, N.J., Awramik, S.M., Petryshyn, V., Spear, J.R., Celestian, A.J., 2019. Neoarchean (2.7 Ga) lacustrine stromatolite deposits in the Hartbeesfontein Basin, Ventersdorp Supergroup, South Africa: Implications for oxygen oases. Precambrian Res. 320, 291–302. https://doi.org/10.1016/j.precamres.2018.11.009

Wogan, N.F., Catling, D.C., Zahnle, K.J., Claire, M.W., 2022. Rapid timescale for an oxic transition during the Great Oxidation Event and the instability of low atmospheric $O_2$. Proc. Natl. Acad. Sci. U. S. A. 119, e2205618119. https://doi.org/10.1073/pnas.2205618119

Zahnle, K.J., Catling, D.C., Claire, M.W., 2013. The rise of oxygen and the hydrogen hourglass. Chem. Geol. 362, 26–34. https://doi.org/10.1016/j.chemgeo.2013.08.004

Zahnle, K.J., Gacesa, M., Catling, D.C., 2019. Strange messenger: A new history of hydrogen on Earth, as told by Xenon. Geochim. Cosmochim. Acta 244, 56–85. https://doi.org/10.1016/j.gca.2018.09.017